# Profiling and Optimizing Java Streams

Eduardo Rosales[a], Matteo Basso[a], Andrea Rosà[a], and Walter Binder[a]

a   Università della Svizzera italiana, Faculty of Informatics, Lugano, Switzerland

**Abstract**
The *Stream API* was added in Java 8 to allow the declarative expression of data-processing logic, typically map-reduce-like data transformations on collections and datasets. The Stream API introduces two key abstractions. The *stream*, which is a sequence of elements available in a data source, and the stream *pipeline*, which contains operations (e.g., `map`, `filter`, `reduce`) that are applied to the elements in the stream upon execution. Streams are getting popular among Java developers as they leverage the conciseness of functional programming and ease the parallelization of data processing.

  Despite the benefits of streams, in comparison to data processing relying on imperative code, streams can introduce significant overheads which are mainly caused by extra object allocations and reclamations, and the use of virtual method calls. As a result, developers need means to study the runtime behavior of streams in the goal of both mitigating such abstraction overheads and optimizing stream processing. Unfortunately, there is a lack of dedicated tools able to dynamically analyze streams to help developers specifically locate issues degrading application performance.

  In this paper, we address the profiling and optimization of streams. We present a novel profiling technique for measuring the computations performed by a stream in terms of elapsed reference cycles, which we use to locate problematic streams with a major impact on application performance. While accuracy is crucial to this end, the inserted instrumentation code causes the execution of extra cycles, which are partially included in the profiles. To mitigate this issue, we estimate and compensate for the extra cycles caused by the inserted instrumentation code.

  We implement our approach in `StreamProf` that, to the best of our knowledge, is the first dedicated stream profiler for the Java Virtual Machine (JVM). With `StreamProf`, we find that cycle profiling is effective to detect problematic streams whose optimization can enable significant performance gains. We also find that the accurate profiling of tasks supporting parallel stream processing allows the diagnosis of load imbalance according to the distribution of stream-related cycles at a thread level.

  We conduct an evaluation on sequential and parallel stream-based workloads that are publicly available in three different sources. The evaluation shows that our profiling technique is efficient and yields accurate profiles. Moreover, we show the actionability of our profiles by guiding stream-related optimizations on two workloads from Renaissance. Our optimizations require the modification of only a few lines of code while achieving speedups up to a factor of 5×.

  Java streams have been extensively studied by recent work, focusing on both how developers are using streams and how to optimize them. Current approaches in the optimization of streams mainly rely on static analysis techniques that overlook runtime information, suffer from important limitations to detect all streams executed by a Java application, or are not suitable for the analysis of parallel streams. Understanding the dynamic behavior of both sequential and parallel stream processing and its impact on application performance is crucial to help users make better decisions while using streams.

## 1 Introduction

Since Java 8, the *Stream API* [1] supports data processing using a declarative style, leveraging the conciseness of functional programming [2]. This API introduces two key abstractions: the *stream*, i.e., a view of a sequence of data elements, and the stream *pipeline*, i.e., a multistage structure of aggregated operations (e.g., map, filter, reduce). The elements on a stream can come from multiple data sources, including collections, arrays, files, generators, and strings. When a stream is executed, the operations in its pipeline are applied to the data elements in the stream [3].

Nowadays, streams are receiving the attention of developers targeting the Java Virtual Machine (JVM) [4–6] as they enable diverse *MapReduce* [7] data transformations [3]. Streams can be used to improve software design by leveraging functional-programming styles that favor both software extensibility and maintainability. Streams can also be executed in parallel by calling a single method, potentially achieving speedups without writing any parallel code [1]. On the other hand, compared to data processing using imperative code, streams can suffer from performance penalties mainly due to extra object allocations and reclamations, as well as the use of many virtual method calls [8, 9].

To mitigate stream-related abstraction overheads and optimize streams, developers need means to study the runtime behavior of sequential and parallel stream executions. To the best of our knowledge, six studies have focused on assisting developers in the optimization of streams. Khatchadourian et al. [10, 11] propose an Eclipse [12] plugin that helps developers better code streams. Møller et al. [9] transform at a bytecode level sequential streams into more efficient imperative code. Basso et al. [13] use bytecode-level transformations to specifically remove overheads due to parallel streams. However, the above approaches mainly rely on the use of static analysis, overlooking dynamic information that is critical to locate stream-related performance issues. Ishizaki et al. [14] and Hayashi et al. [15] exploit GPUs to speed up stream processing. However, both approaches can be applied only to a single method of the Stream API and require a proprietary compiler, reducing their applicability.

While several profilers for the JVM enable multiple analyses and target many metrics [16–21], there is no profiler able to specifically capture the runtime behavior of all sequential and parallel streams executed by a Java application. Overall, there is a lack of dedicated tools able to dynamically analyze streams, which help developers locate streams with a significant impact on application performance, and provide *actionable profiles*[1] guiding stream-related optimizations. We aim at filling this gap, addressing the dynamic analysis of streams to help developers locate stream-related performance drawbacks.

**Contributions.** We propose a novel technique for profiling all the streams executed by an application on the JVM (Sec. 2). We measure the computations performed by a stream in terms of *reference cycles*, defined as the clock cycles elapsed during

---

[1] Mytkowicz et al. [22] define a profile as *actionable* if acting on it yields the expected outcome, e.g., acting on the code portions indicated by a profile results in better performance.

Eduardo Rosales, Matteo Basso, Andrea Rosà, and Walter Binder

an operation, collected at the nominal processor frequency (regardless of frequency scaling). Profiling the reference cycles (henceforth called *cycles* for short) elapsed by stream processing serves multiple purposes. First, it allows identifying expensive stream executions, whose optimization may yield speedups. Second, it allows the developer to better assess the benefits of possible stream parallelization (considering whether sequential stream computations are expensive enough to benefit from a parallel execution). Third, it allows the detection of load imbalance affecting parallel stream processing. Finally, it helps detect abundant small streams executing only a few cycles (e.g., created via flatMap [23]), which can give developers clues in spotting optimizations (e.g., removing such small streams by using mapMulti [23]).

While accuracy is crucial to our aim, the inserted instrumentation code enabling cycle profiling inevitably causes the execution of extra (undesirable) cycles, which are (partially) included in the profiles. To mitigate this problem, we *compensate* for the extra cycles caused by the instrumentation, i.e., such cycles are estimated and removed from the collected profiles.

We implement our approach in StreamProf that, to the best our knowledge, is the first dedicated stream profiler for the JVM. We demonstrate the usefulness of StreamProf by profiling two workloads in the state-of-the-art benchmark suite Renaissance [24], diagnosing and solving stream-related performance issues (Sec. 3). We first locate unneeded streams that can be safely removed without altering the semantics of the workloads. We then diagnose load imbalance affecting parallel stream processing. We use the actionable profiles produced by StreamProf to fix the detected problems, achieving speedups up to a factor of 5×. We are not aware of any other study targeting the optimization of stream-based applications in the Renaissance suite.

Through an evaluation conducted on workloads that are publicly available in three different sources, we measure the accuracy of the profiles (after compensation) produced by StreamProf and the profiling overhead (Sec. 4). Our evaluation results show that our approach is efficient and produces accurate, actionable profiles that help developers optimize stream processing, while introducing a profiling overhead that enables its practical use in Java applications.

Our work faces multiple challenges. Differently from related approaches [9–11, 14, 15], our technique targets the detection of every sequential and parallel stream executed by a Java application. As a result, a first challenge arises from developing an instrumentation properly tracking every form of stream execution available in the Java Class Library (JCL). Another key challenge we address is the need for reducing the profiling overhead to enable cycle-accurate stream profiling. This requires an efficient instrumentation collecting a meaningful and minimal selection of metrics to reduce measurement perturbation. Moreover, we address the estimation of the extra cycles caused by the inserted instrumentation code, to attempt removing them from the profiles via a perturbation-compensation technique.

We complement the paper with a discussion of limitations of our approach and threats to validity in our evaluation (Sec. 5), a review of related work (Sec. 6), and concluding remarks (Sec. 7).

## 2 StreamProf

Here, we describe our approach to profile streams on the JVM. We first present the profiling model, which describes the key entities and events targeted by our technique (Sec. 2.1). Then, we detail our approach to profile stream executions (Sec. 2.2).

### 2.1 Profiling Model

Listing 1 shows a Java code example of a stream. In line 2, a stream is generated from bookList, i.e., a list of objects of class Book. The pipeline has four operations: filter that selects top-seller books (line 3), flatMap which *flattens*[2] the list of authors of each book in bookList (line 4), and collect that accumulates top-seller author names into a list (line 5). We use this example to describe our profiling model in this subsection.

■ **Listing 1** A simple Java code stream example.

```
1 List<String> getTopSellerAuthors(List<Book> bookList) {
2   return bookList.stream()
3     .filter(Book::isTopSeller)
4     .flatMap(b -> b.getAuthors().stream())
5     .collect(Collectors.toList());
6 }
```

```
1 class Book {
2   ...
3   boolean isTopSeller() {...}
4   List<String> getAuthors() {...}
5   ...
6 }
```

**Stream.** As a *stream*, we consider every instance of a subtype of java.util.stream.AbstractPipeline, the abstract class representing a stream and the operations in its pipeline. As any form of stream creation in the JCL ends up instantiating a subtype of AbstractPipeline, we instrument all these subtypes to guarantee the detection of every stream execution regardless of how the stream was created. In the Stream API, one can create a stream which data elements are either objects or values of primitive types (i.e., only int, long, or double).

**Stream execution.** Our profiling technique detects *stream executions*. A stream is executed by calling a *terminal operation*, i.e., an operation triggering the *traversal* of the pipeline either to return a result (e.g., an array) or to produce side effects (e.g., to print all elements). Streams are *lazy*, i.e., they do not perform any processing until a terminal operation is invoked. A stream can have at most one terminal operation, which can be executed only once [1]. In listing 1, collect is the terminal operation. A terminal operation invokes a *stream execution method*, i.e., a method in the Stream API (located either in AbstractPipeline or its subtypes) that begins the execution of a stream. Our approach supports *infinite streams* (i.e., a stream which data source is capable of generating an infinite number of elements), measuring the reference cycles elapsed during the execution of the stream until it is completed. Also, streams can be

---
[2] The term *flattening* refers to merging elements contained in multiple collections/arrays into a single data structure.

Eduardo Rosales, Matteo Basso, Andrea Rosà, and Walter Bindereither *sequential* or *parallel*. The former are synchronously executed by the current thread, while the execution of the latter involves the creation of tasks.

**Tasks.** When a parallel stream is executed, its computations are carried out by *fork/join*[3] *tasks*. An initial task is spawned upon calling the terminal operation, forking more tasks if needed. As tasks, we need to consider all subtypes of java.util.concurrent.-CountedCompleter in the implementation of the Stream API. Tasks are executed in a common *fork/join pool* [26] by threads called *workers*. The fork/join pool implements *work stealing* [27, 28]. In the Stream API, a worker executes a task by invoking an implementation of the method CountedCompleter.compute,[4] to which we refer as *task execution method*. We consider a task to be *executed* after its execution method has completed. We profile all tasks spawned by parallel streams.

Pipelines containing *stateful operations* (i.e., intermediate operations where the current state may depend on the state of previously seen elements) may require to buffer significant data, potentially impairing parallel performance [1]. Our approach partially measures data buffering affecting parallel streams since we only measure the computations performed by the tasks spawned to execute the stream, disregarding thread-scheduling overheads in the OS (see the limitations of our technique in Sec. 5).

**Nested streams.** A stream can be *nested*, i.e., its execution is triggered by (and occurs during the execution of) another stream, i.e., the *outer stream*. The execution of a nested stream always completes before its outer stream. In listing 1, the flatMap operation uses streams that are nested to the outer stream created from bookList. An outer stream can also be nested, thus multiple *nesting levels* are possible. Stream-nesting information can help developers better assess the benefits of possible stream parallelization, e.g., sequential streams nested in a parallel stream may not benefit from further parallelization. Also, detecting *inefficient nesting*, e.g., abundant nested streams where each one only runs for a few cycles, may indicate optimization opportunities, as such streams could be merged or avoided to improve application performance. Concretely, mapMulti [23] operations can use imperative code that can avoid inefficient nesting otherwise introduced by operations such as flatMap. Our approach properly reconstructs the nesting relationships between streams.

**Reference cycles.** Compared to other metrics, e.g., *wall time* [29] and *bytecode count* [30], reference cycles provide a precise measure of the computations performed by a stream, excluding intervals where a worker was not scheduled to execute on a CPU core and accounting for latencies due to misalignments and cache misses. Complete cycle profiling of all stream executions helps detect stream-related performance issues (as shown in Sec. 3).

**Location.** We associate stream executions with a *location*, i.e., the fully qualified name of the caller of the terminal operation. This information shows developers the class and method where each stream is executed in application code. In listing 1, the

---

[3] Fork/join parallelism relies on the recursive splitting (*fork*) of work into tasks that are executed in parallel, waiting for them to complete, and then typically merging (*join*) the results computed by the forked tasks [25].
[4] The fully qualified name of a class/interface appears upon first occurrence in the text; thereafter, we use only the class/interface name.

10:5

**Profiling and Optimizing Java Streams**

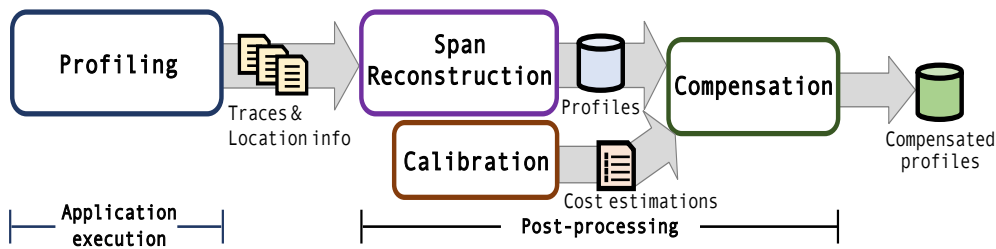

**Figure 1** Overview of the profiling methodology.

location of the outer stream is method getTopSellerAuthors (omitting the class name). Locations are a key part of an actionable profile, directing the developer to application code where optimizations may have the biggest performance impact.

**Span.** We model stream execution around the concept of *span*, which we define as the interval in which a stream is executed by a thread. The beginning and the end of each span are defined by reference-cycle readings just before/after stream execution. Our goal is to profile all spans during the execution of a target application.

To collect the cycles elapsed in the span, we read a per-thread virtualized reference-cycle counter provided by the processor's Performance Monitoring Unit. We call $cycles_{begin}$ and $cycles_{end}$ the values in cycles read at the beginning and end of the span, respectively. We define the quantity $cycles_{end} - cycles_{begin}$ as the *measured cycles* for the span. As each thread-cycle counter is strictly monotonically increasing in terms of elapsed cycles, we assume that $cycles_{begin}$ and $cycles_{end}$ are totally ordered within the scope of a single thread. Moreover, spans in a thread cannot overlap, but a span can be within another one, enabling the detection of nested streams.

A span always refers to a single stream, but multiple spans can be associated with the same parallel stream. A sequential stream execution is represented by a single span which is defined by reference-cycle readings just before and after the stream execution method. In contrast, a parallel stream execution is in general represented by multiple spans (i.e., one per task spawned to carry out the parallel stream computation), where each span is defined by reference-cycle readings just before and after the task execution method. The computations carried out by each task of the same parallel stream contribute to the measured cycles of the stream.

## 2.2 Profiling Streams on the JVM

Fig. 1 shows our approach to stream profiling. First, StreamProf instruments the target application to enable the *profiling* of all spans detected at runtime. StreamProf builds on DiSL [31], a dynamic analysis framework for the JVM based on bytecode instrumentation. A key aspect of DiSL is that it enables the complete instrumentation of the JCL, in which the Stream API is implemented. Thanks to DiSL, StreamProf can accurately intercept every stream executed by a Java application.

At the end of application execution, StreamProf produces one trace for each thread that executed at least one stream. Each trace contains information on all the spans representing stream executions performed by the corresponding thread. In addition to per-thread traces, StreamProf stores the locations in a separate file (referenced by the traces via location indices). This file and the traces are processed offline, to minimize data processing while profiling, thus reducing measurement perturbation. During post-processing, we first perform *span reconstruction*, i.e., we produce per-stream profiles considering all the per-thread traces. The information in each stream profile is reconstructed from the span(s) related to the corresponding stream.

We also perform a *calibration* phase (only once per profiling platform), to obtain estimated costs, i.e., estimations of the extra cycles required to profile spans. Such cost estimations are used in the last phase, where we *compensate* for the extra cycles due to the instrumentation, i.e., we remove them from the measured cycles of each span. The output of StreamProf are profiles reporting the compensated cycles, nesting relationships, and the location of each stream execution, including also aggregated statistics at a thread- and location-level (see Sec. 3). Overall, our profiles report accurate information about all streams executed by a target application. In the rest of the section, we describe each phase in more detail.

### 2.2.1 Profiling

The aim of the profiling phase is intercepting the beginning and end of every span to associate it with the corresponding measured cycles and location, storing the collected data in memory during application execution and persisting that information upon application completion.[5]

**Spans.** A sequential stream execution is carried out by a single thread and is represented by a single span. This enables directly attributing the measured cycles in the span to the corresponding stream, without requiring any ID to further identify the sequential stream. For this reason, we use the term *anonymous span* to refer to the single span of a sequential stream execution.

The execution of a parallel stream is generally performed in multiple tasks that can be executed by several workers, thus requiring the detection of multiple spans that can relate to different threads. Differently from sequential streams, in this case it is necessary to attribute all spans to the corresponding parallel stream, such that the measured cycles in each span can be aggregated to estimate the whole computations performed by the stream. By design of the Stream API, parallel stream execution always starts with the execution of a primordial fork/join task that can spawn other tasks if needed. When the primordial task is executed, we associate a unique ID with the corresponding parallel stream, i.e., we associate the stream ID with the pipeline *head*, the object representing the initial stage of the pipeline. When profiling the span corresponding to a forked task of the same parallel stream, the instrumentation associates the span with the stream ID (i.e., as tasks have a reference to the head of the pipeline, we use it to query the stream ID). Hence, we use the term *named spans* to refer to spans related to a parallel stream execution.

Among all the named spans associated with the same stream ID, exactly one span corresponds to the execution of the stream's primordial task. We refer to that span as the *primordial span*, while we use the term *support span* to denote any additional named span of the same stream. When profiling a primordial span, there is always the need for generating a stream ID, whereas support-span profiling requires only retrieving an existing stream ID. As explained in Sec. 2.2.2, primordial spans also allow identifying the outer stream (if any), which is required to reconstruct stream nesting. Lastly, we remark that our technique detects every stream execution, regardless of the thread executing the stream or the fork/join pool configuration. Moreover, our technique enables the profiling of all fork/join tasks of parallel stream executions while disregarding any other processing concurrently using the fork/join pool.

---

[5] The instrumentation logic in pseudocode is presented in Appendix A.

**Location.** To attribute a location to each span, we perform a dedicated analysis (upon instrumentation) to determine the fully qualified name of all methods that may directly invoke stream execution methods and identify them with a unique method ID. Upon intercepting the beginning of a span, the instrumentation recognizes the caller of the stream execution method, retrieves its pre-computed method ID, and associates it with the current span. A file containing the mapping between all method IDs and the corresponding method names is dumped upon application completion.

**Trace generation and dumping.** The instrumentation relies on a *tracer*, i.e., a JVMTI [32] agent attached to the target application, in charge of reading cycle counters from native code, storing span information, and dumping traces. Upon span begin, the instrumentation uses the tracer both to read $cycles_{begin}$ and to store it along with the location in the form of an event (*span-begin*). Similarly, the instrumentation uses the tracer to read $cycles_{end}$, storing the value in a second event (*span-end*). The tracer resorts to *PAPI* [33] (i.e., an open-source C library providing a common interface to read low-level per-thread virtualized counters) to read the event *PAPI_REF_CYC*.

To reduce profiling overheads, the tracer maintains thread-local buffers where all span information associated with a thread is stored in memory, eliminating the need for synchronization (apart from when allocating new buffers). Upon application completion, the tracer dumps the buffers, generating one trace per thread. When the pre-allocated buffers are full, the tracer dumps them and allocates new buffers at runtime. We remark that none of the analyzed workloads analyzed in Sec. 3 and Sec. 4 required this. Lastly, each dumped trace contains the ID and the name of the thread, as well as the observed sequence of *span-begin* and *span-end* events. Overall, to mitigate measurement perturbations, we store the least possible amount of data.

### 2.2.2 Span Reconstruction

We process the traces to compute per-stream information, namely the total measured cycles in each stream and the location.[6] We first process each trace individually, generating a profile for each individual thread. The profile of a stream is reconstructed from its respective *span-begin* and *span-end* events that are stored in the traces. To this end, we maintain a stack for each thread where for each *span-begin* event, a new span object is pushed onto the stack, while the respective *span-end* event closes the processing of the span, popping it off the stack.

---

[6] The span reconstruction algorithm in pseudocode is shown in Appendix B.

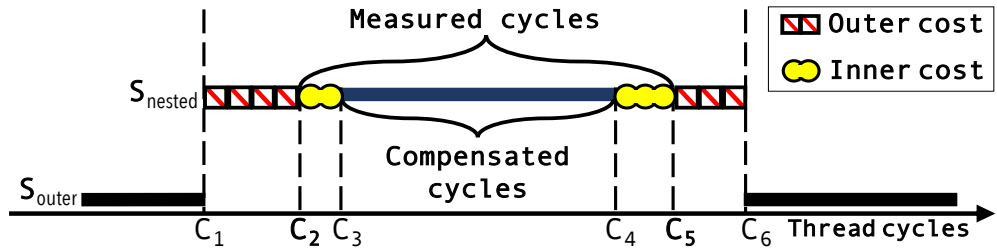

**Figure 2** Representation of inner and outer costs. $C_{\#}$ represents a value in cycles as read at a given time from the thread-cycle counter. $C_2$ represents $S_{nested_{begin}}$ and $C_5$ represents $S_{nested_{end}}$.

For sequential stream execution, the measured cycles are computed from the respective single anonymous span. In contrast, the named spans of a single parallel execution are typically in multiple thread profiles. Therefore, we need to merge the thread profiles, associating each named span with the parallel stream execution whose ID was attributed to the span by the instrumentation. After merging, the measured cycles of a parallel stream execution are computed by aggregating the measured cycles in all its named spans. Each span is also assigned its corresponding location. To do so, we look up the span's method ID in the location-mapping file. For a parallel stream, all its named spans share the same location.

The stack also enables reconstructing the nesting relationships between spans. For an anonymous span, the stack allows identifying the corresponding outer span, i.e., the span just below the top of the stack, if any. For named spans of the same parallel stream execution, the stack provides valid outer-span information only for the primordial span, as eventual support spans represent tasks that may have been executed by different threads. Therefore, for each parallel stream execution, we use its primordial span as the representative, such that support spans inherit the outer-span information from it. Once outer-span information is updated, we compute the nesting level for each span. A span is assigned the nesting level zero if it has no outer span, otherwise, the nesting level is recursively computed for the chain of outer spans until the outermost span is found.

### 2.2.3 Calibration

The amount of measured cycles in a span is a key metric to quantify the computations performed by a stream. Still, the instrumentation code inserted by `StreamProf` causes the execution of extra cycles that are (partially) included in the measured cycles of each span, decreasing profile accuracy. We aim at removing these extra cycles from the generated profiles, such that the cycles attributed to each span better match those of an uninstrumented application execution. To do so, the extra cycles need to be estimated, which is the aim of the calibration phase.

**Cost of instrumentation.** Henceforth, we use the term *cost* to refer to the extra cycles introduced by the instrumentation to profile a span. We estimate two types of cost as illustrated in Fig. 2. The figure shows the elapsed cycles associated with a span $S_{nested}$ that is nested in an outer span $S_{outer}$. The x-axis represents the values of the thread cycle counter, which increases as stream execution proceeds.

**Profiling and Optimizing Java Streams**

When profiling $S_{nested}$, StreamProf executes instrumentation code to profile its begin and end. Suppose that the instrumentation code to profile the begin of $S_{nested}$ takes $C_3$–$C_1$ cycles (i.e., the measure of cycles elapsed between $C_1$ and $C_3$), whereas the one to profile the end of $S_{nested}$ takes $C_6$–$C_4$ cycles. While profiling such events, StreamProf reads the cycle counter twice to obtain $cycles_{begin}$ and $cycles_{end}$. Suppose that $cycles_{begin}$ and $cycles_{end}$ are the cycle values $C_2$ and $C_5$, respectively. The real amount of cycles elapsed during the execution of $S_{nested}$ is $C_4$–$C_3$. However, the cycles measured by the profiler are $C_5$–$C_2$, which is greater than the real amount of cycles. As *inner cost*, we define the quantity $(C_3$–$C_2) + (C_5$–$C_4)$, which represents the cost accounted within the measured cycles of the span. Note that each span is subjected to the inner cost.

When the profiler computes the cycles of $S_{outer}$, it subtracts the measured cycles of $S_{nested}$ (i.e., $C_5$–$C_2$). When doing so, as the measured cycles in $S_{nested}$ include its inner cost, the inner cost of $S_{nested}$ is not accounted in $S_{outer}$. However, the instrumentation of $S_{nested}$ also causes the execution of extra cycles before reading $cycles_{begin}$ ($C_2$–$C_1$) and after reading $cycles_{end}$ ($C_6$–$C_5$). These cycles are not accounted in $S_{nested}$, but are accounted in $S_{outer}$. As *outer cost*, we define the quantity $(C_2$–$C_1) + (C_6$–$C_5)$, corresponding to the extra cycles measured only in the outer span.

The outer cost depends on the kind of span. The lowest outer cost occurs when profiling anonymous spans, as no stream ID is needed. For named spans, the outer cost depends on whether a stream ID has to be generated (i.e., in primordial spans) or only retrieved (i.e., in support spans). Primordial spans have a higher outer cost than support spans, due to the computations required to generate a new stream ID.

We remark that Fig. 2 shows that our instrumentation costs are generally smaller in scale than the measured cycles of a stream execution. However, the extra cycles introduced by the execution of instrumentation code can be dominant in some cases, e.g., when executing nested streams running only for a few cycles.

**Cost estimation.** We aim at a simple estimation of the inner and outer costs, approximating them with constants. We denote as *IC*, *OC_ANON*, *OC_PRIM*, and *OC_SUPP*, respectively, an approximation of the inner cost, the outer cost of an anonymous span, of a primordial span, and of a support span. The calibration phase measures these constants for a given platform. To this end, we use a *span generator*, i.e., a Java program that generates spans arranged in pairs, one nested within the other. No code is executed during the nested span, while during the outer span we execute code that corresponds to the span-type-dependent begin/end-events of the nested span.

To estimate *OC_ANON*, during the outer span only the code to profile the begin/end-events of the nested anonymous span is executed. To estimate *OC_PRIM*, during the outer span we additionally execute the computations required to generate a new stream ID. To estimate *OC_SUPP*, during the outer span we execute the computations required to retrieve an already existing stream ID. In this calibration setting, the measured cycles for the nested span correspond to *IC*; this can be seen in Fig. 2, assuming that the compensated cycles are zero (because during the nested span no code is executed). We estimate *IC* as the average of the measured cycles for the nested spans. Then, we estimate the different outer costs as shown in Eq. 1.

$$outer\_cost = (S_{nested.cycles_{begin}} - S_{outer.cycles_{begin}}) + (S_{outer.cycles_{end}} - S_{nested.cycles_{end}}) - IC \quad (1)$$

Note that *outer_cost* is a generic term that can refer to any of the three types of outer cost. The actually computed cost depends on the kind of span generated (e.g., if the generated span is anonymous, then *OC_ANON* would be computed). Each cost is estimated as the average obtained from the execution of 10 million pairs of spans, from which outliers are removed.

#### 2.2.4 Compensation

The goal of the compensation phase is to compute *compensated cycles* for each span as shown in Eq. 2.

$$compensated\_cycles = (measured\_cycles - nested\_cycles) - \qquad (2)$$
$$(n_{anon} \cdot OC\_ANON) - (n_{prim} \cdot OC\_PRIM) - (n_{supp} \cdot OC\_SUPP) - IC$$

We first subtract *nested cycles* (i.e., the cycles elapsed due to nested spans) from the measured cycles (cf. Fig. 2) of each outer span. This is necessary to account only for the cycles of an outer span as nested cycles are included by default in the outer-span measured cycles. To enable this, for each span we maintain information about the directly nested spans, including the total nested cycles. Then, all costs (approximated by constants) are also subtracted. We do so to remove an estimation of the cycles caused by the execution of the inserted instrumentation code. To this end, for each span, we maintain the total number of nested anonymous ($n_{anon}$), primordial ($n_{prim}$), and support ($n_{supp}$) spans, which are multiplied by the respective estimated outer cost. Lastly, we subtract the inner cost to which each span is subjected (see Sec. 2.2.3).

## 3 Profiling and Optimizing Streams

In this section, we show how StreamProf helps developers optimize streams. We first describe our experimental setup (Sec. 3.1). Then, we present the optimization of two workloads from the Renaissance benchmark suite, first locating and removing unneeded stream executions (Sec. 3.2), and then improving load balance (Sec. 3.3). Finally, we show the speedups obtained thanks to our optimizations (Sec. 3.4).

### 3.1 Experimental Setup

In this subsection, we describe our experimental setup.

**Benchmarks.** Our evaluation targets Renaissance [24], a state-of-the-art benchmark suite for the JVM with modern, concurrent, object-oriented workloads. In contrast to previous well-known suites, such as DaCapo [34], ScalaBench [35], and SPECjvm2008 [36], Renaissance includes workloads heavily exercising stream processing. We use the latest release of Renaissance at the time of writing, i.e., v.0.14.1, released on May 23, 2022.

Renaissance includes three workloads exercising streams, two of which we optimize, i.e., mnemonics and its parallel version, par-mnemonics. We remark that the third

**Profiling and Optimizing Java Streams**

stream-based workload from Renaissance, i.e., scrabble, is considered only in our evaluation (Sec. 4) as we found no stream-related performance issue to be optimized in the workload. For our experiments, we consider only the *steady state*, i.e., the state in which garbage-collector ergonomics and dynamic compilation are stable after running enough *warm-up* iterations. For both mnemonics and par-mnemonics, we run 16 warm-up iterations as suggested by the developers of Renaissance [37].

**Testbed.** We run all the experiments on two machines $M_1$ and $M_2$. $M_1$ has an 8-core Intel Xeon E5-2680 (2.7 GHz) and 128 GB of RAM. $M_2$ has an 18-core Intel i9-10980XE (3.0 GHz) with 256 GB of RAM. $M_1$ and $M_2$ run under Linux, generic kernel versions 4.15.0-147 and 5.4.0-89, respectively. We disable Turbo Boost and Hyper-Threading. The CPU governor is set to "performance." We use Oracle JDK 17 [38] (build 17.0.3.1) and PAPI 6 (build 6.0.0.1).

### 3.2 Optimization 1: Removing Unneeded Streams

We first address the profiling and optimization of mnemonics. StreamProf produces *customized heatmaps*[7] that can be used by the developer to understand stream nesting as well as the distribution of stream executions and compensated cycles. In our approach, we focus on the darker regions indicated by the heatmap, as optimizations that reduce the cycles in these regions are likely to improve performance. To enable this task, StreamProf also reports *hot locations*, i.e., locations in application code responsible for most of the stream processing. Thanks to this information, we first analyze the *hottest location*, meaning the location with both the highest number of stream executions and total cycles, i.e., org.renaissance.jdk.streams.MnemonicsCoderWithStream.wordForNum. The streams executed in this method stand for 90.07% of the total streams (14, 532, 738) executed by mnemonics, accounting for 76.43% of its total cycles.

We inspect the sources and find that method wordForNum is executed as part of a recursive processing. As recursive calls take place as part of stream execution, this explains the deep nesting detected by our profiler. We also find that stream executions triggered from method wordForNum always derive from the same data source, i.e., a list of strings which is initialized before workload execution and remains unchanged thereafter. Moreover, we find that the execution of such streams does not depend on any input of the current recursion level, thus they always produce the same output. To optimize mnemonics, these streams can be moved out of the recursive code and executed only once.

Following the same strategy, we identify another hot location, i.e., MnemonicsCoderWithStream.lambda$encode$9. We find that the stream executions at this location account for 3.75% of the total streams executed by mnemonics, yet accounting for 10.64% of its total cycles. We find that some of these streams are created from a set passed as an argument to method MnemonicsCoderWithStream.encode. Furthermore, this set is initially empty and is never modified. Moreover, we check that the output produced after the execution of the streams created from the empty set does not contribute in any form to the output of mnemonics, showing that we can safely remove these unneeded stream executions.

---

[7] The heatmaps produced by StreamProf are shown in Appendix C.

We apply both modifications to mnemonics, removing the problematic streams detected and notably reducing unneeded and expensive computations in the optimized version.[8] Overall, the reduction in stream executions significantly improves performance (see Sec. 3.4). We aim at optimizations with only minimal changes to the original workload. Thanks to our actionable profiles, we optimized mnemonics by changing only 6 lines of code.

### 3.3 Optimization 2: Improving Load Balance

We optimize par-mnemonics following the same approach described in Sec. 3.2, targeting methods MnemonicsCoder.paralleEncode and MnemonicsCoder.wordsForNum-Parallel, which represent the counterpart of the methods optimized in mnemonics, triggering parallel (instead of sequential) stream processing. In the profiles produced for the optimized par-mnemonics, we run an extra analysis provided by StreamProf to compute *load balancing information*, i.e., the distribution of cycles per worker involved in parallel stream processing. We use the *coefficient of variation*[9] (CV) to analyze the variability of stream-related per-worker cycles. A CV equal to zero would indicate a perfect load balance while the higher the CV, the higher the imbalance.

We find that par-mnemonics suffers from poor load balance. Despite 8 workers carry out the execution of par-mnemonics on $M_1$, only two of them execute 99.99% (79.54% and 20.45%, respectively) of the total cycles, leading to a CV of 2.24. Similarly, 10 workers execute par-mnemonics on $M_2$, where two of them execute 99.99% (80.67% and 19.32%, respectively) of the total cycles, for a CV of 2.55. On both machines, 52 tasks are spawned in total. Our profiles indicate the location where parallel stream execution is first triggered in the workload, i.e., MnemonicsCoderWithStream.parallelEncode. We find that only the first recursion level performed by par-mnemonics executes parallel streams. In contrast, subsequent recursion levels rely on sequential streams. This behavior is defined via a threshold that considers the number of elements to be processed at the current recursion level.

On a fork/join pool, a key strategy to improve load balance is increasing the number of tasks spawned, which can keep more CPU cores busy, improve scalability and locality, and reduce the time that CPU cores are idle [39]. However, spawning an excessive number of tasks may overshadow the benefits of parallel processing due to overheads introduced by the creation, scheduling, and managing of too many tasks [40]. To reduce the load imbalance of par-mnemonics, we optimize task granularity.

We increase the aforementioned threshold such that more recursion levels use parallel streams. We obtain the best results with a threshold of 20 (the original one is 3), resulting in $226,596$ tasks spawned on both $M_1$ and $M_2$. The increased number of tasks results in an improved CV of 0.60 and 0.57 on $M_1$ and $M_2$, respectively.

---

[8] The heatmap after our optimizations is shown in Appendix C.
[9] The coefficient of variation is the ratio of the standard deviation to the mean.

■ **Table 1** Speedups due to optimizing stream processing.

| Bench. | Ver. | Machine | #Streams | #Tasks | Time [ms] | Speedup Factor | 95% CI |
|---|---|---|---|---|---|---|---|
| mne | orig | $M_1$ | 14,532,738 | | 4,944.72 | | |
| | | $M_2$ | | | 3,303.64 | | |
| | opt1 | $M_1$ | 1,088,946 | | 1,200.70 | **4.09** | (3.99, 4.19) |
| | | $M_2$ | | | 981.97 | **3.36** | (3.32, 3.40) |
| par-mne | orig | $M_1$ | 14,532,738 | 52 | 4,419.53 | | |
| | | $M_2$ | | | 2,977.83 | | |
| | opt1 | $M_1$ | 1,088,946 | 52 | 1,106.22 | **3.98** | (3.89, 4.04) |
| | | $M_2$ | | | 905.19 | **3.27** | (3.20, 3.33) |
| | opt2 | $M_1$ | 1,088,946 | 226,596 | 880.09 | **5.00** | (4.91, 5.10) |
| | | $M_2$ | | | 764.63 | **3.88** | (3.80, 3.96) |

To sum up, we optimized load balance in par-mnemonics thanks to our actionable profiles. Our optimization requires changing only 2 lines of code (in addition to the 6 lines reported in Sec. 3.2). We note that achieving a better CV would likely require a deep code refactoring of par-mnemonics, which is beyond the scope of our work.

### 3.4 Speedups

In this subsection, we report the speedups enabled by our optimizations.

Table 1 summarizes information on the stream processing in both the original (*orig*) and optimized versions (*opt1* and *opt2*), discriminated in column *Ver.*, of mnemonics (mne) and par-mnemonics (par-mne) on $M_1$ and $M_2$. First, we report the total amount of streams executions (*#Streams*), the total number of tasks (*#Tasks*, only in par-mnemonics), and the workload execution times (*Time*). Then, we present the *speedups* [10] achieved by our optimizations. The execution times and speedups reported are the average [11] of 100 steady-state runs. We report 95% *confidence intervals* (CI) for each speedup. All optimized workloads pass the validation performed by the harness of Renaissance, indicating that our modifications do not alter workload output.[12]

For all modified workloads, we observe remarkable speedups thanks to our optimizations. First, removing unneeded stream processing (*opt1*) leads to significant speedups in both mnemonics (up to 4.09× on $M_1$ and 3.36× on $M_2$) and par-mnemonics (up to 3.98× on $M_1$ and 3.27× on $M_2$). Second, in addition to the removal of unneeded streams, improving load balance (*opt2*) results in an important performance gain in par-mnemonics (up to 5.00× on $M_1$ and 3.88× on $M_2$).

Overall, cycle profiling is effective to detect problematic streams (among the millions executed) whose optimization has the potential to improve application performance. Accurately profiling fork/join tasks supports the diagnosis of load imbalance according to the distribution of stream-related cycles at a worker level. Moreover, the actionable profiles generated by StreamProf help developers focus on the code locations where problematic streams are executed. Indeed, performing each optimization (i.e., manual code inspection, code changes, and thorough testing) took us around two hours.

---

[10] A speedup is calculated as the execution time of the original (unmodified) workload divided by the execution time of the optimized workload.
[11] Averages across multiple workloads are computed using the geometric mean.
[12] We have contacted the developers of Renaissance, who confirmed both the correctness and benefits of our optimizations.

## 4 Evaluation

In this section, we present the evaluation of StreamProf. We first describe our experimental setup (Sec. 4.1). Then, we evaluate the accuracy of the profiles produced by StreamProf (Sec. 4.2) and its profiling overhead (Sec. 4.3).

### 4.1 Evaluation Setting

Our evaluation targets workloads from three publicly available sources. First, we target all workloads studied in Sec. 3, including also scrabble, a parallel workload executing 4,350,761 streams and spawning 32 and 128 tasks on $M_1$ and $M_2$, respectively. By including scrabble, we target all the stream-based workloads available in Renaissance. Second, we target workloads from JEDI [41, 42], a benchmark suite consisting of 25 stream-based applications, each recasting a *TPC-H* [43] query into a Java implementation making use of the Java Stream API. Lastly, we target selected[13] stream-based workloads written by the developers of the OpenJDK and publicly available in the OpenJDK18 open-source repository [44] (we refer to these workloads as OpenJDK for short). We run 50 warm-up iterations for scrabble (as suggested by the developers of Renaissance [37]) using the harness of the suite, while we run 15 warm-up iterations for both the workloads from JEDI and OpenJDK using the *Java Microbenchmark Harness* (JMH) [45]. We perform our evaluation on the same testbed described in Sec. 3.1. We remark that we used StreamProf to analyze the JEDI and OpenJDK workloads and found no stream-related performance issue to be optimized.

We identify two relevant dimensions of interest for our evaluation. The first dimension considers the average *cycles per span* (CPS). We aim at understanding the impact of different CPS on accuracy and profiling overhead. We expect that for workloads with a low CPS, the relative cost of the inserted instrumentation code (in proportion to the cost of stream processing) is higher than for workloads with a higher CPS. Hence, we expect that a higher CPS results in both higher accuracy and lower overhead. The second dimension considers the *execution mode* of stream processing, i.e., sequential vs. parallel. We aim to study whether the accuracy of our profiles or the profiling overhead are affected in the presence of parallel stream execution.

### 4.2 Accuracy

In this subsection, we evaluate the accuracy of the profiles generated by StreamProf.

---

[13] Since OpenJDK includes some very similar workloads (e.g., using method references instead of lambdas), we consider only one representative for each kind of workload to avoid bias in our evaluation.

**Table 2** Estimated costs of the instrumentation in cycles. *IC, OC_ANON, OC_PRIM,* and *OC_SUPP*, respectively denote the inner cost, the outer cost of an anonymous span, of a primordial span, and of a support span.

| Cost | $M_1$ AVG | CV | $M_2$ AVG | CV |
|---|---|---|---|---|
| *IC* | **171.63** | 0.24 | **130.12** | 0.32 |
| *OC_ANON* | **184.25** | 0.24 | **132.46** | 0.41 |
| *OC_PRIM* | **212.40** | 0.21 | **169.82** | 0.36 |
| *OC_SUPP* | **201.17** | 0.23 | **163.83** | 0.44 |

#### 4.2.1 Costs of Instrumentation

As explained in Sec. 2, we conduct a calibration phase to estimate the instrumentation costs as constants. Table 2 summarizes the constants computed for the inner and the different types of outer cost on both $M_1$ and $M_2$. Following the cost estimation method reported in Sec. 2.2.3, we generate 10 million pairs of spans per cost on both machines. The table reports the average costs obtained on these spans, along with the CV.

The CV on $M_2$ is higher than the one on $M_1$. Since our costs are estimated in scenarios where named spans are used, an increased variability is expected because $M_2$ has more cores ($M_1$ is an 8-core and $M_2$ is an 18-core machine).

One goal of our approach is mitigating the perturbations due to the execution of instrumentation code. The low values reported in Table 2 are the result of using an efficient instrumentation collecting minimal data, along with fine-tuned native code (i.e., the tracer) enabling the efficient collection of cycles and their temporal storage in memory during application execution.

#### 4.2.2 Baseline Computation

The goal of our accuracy evaluation is comparing the total compensated cycles accounted for all the profiled spans of a target application as reported by StreamProf with the stream-related cycles elapsed by the same application in the absence of any profiling, i.e., the *baseline*. The smaller the difference between the total compensated cycles and the baseline, the better the accuracy. We focus on the total compensated cycles across all the profiled spans and not on individual spans, because we are not aware of any method to obtain a reliable baseline at a single-span level. We discuss threats to validity in our evaluation in Sec. 5.

Computing a precise baseline is crucial for our evaluation. However, the baseline can only be determined via a measurement, which in turn is subjected to perturbations. To reduce perturbations, we take advantage of the fact that the computations performed by all evaluated workloads take place only within stream executions. Therefore, we measure the total cycles elapsed during the execution of a workload (which we approximate to the total cycles elapsed in stream-related computations) and take them as the baseline.

To compute the baseline, we use PAPI to read cycle counters only twice per thread, i.e., before and after the thread carries out stream processing during the steady-state iteration. For workloads executing sequential streams, we attach PAPI only to the thread triggering workload execution, i.e., the *main thread*. For workloads executing

■ **Table 3** Results of the accuracy evaluation.

| Bench. | Ver. | Machine | #Anon. Spans | #Prim. Spans | #Supp. Spans | Total cycles AVG | CV | CPS | Accuracy [%] |
|---|---|---|---|---|---|---|---|---|---|
| mne | orig | $M_1$ | 14,532,738 | | | 13,304,591,663 | 0.13 | 915 | **87.99** |
| | | $M_2$ | | | | 9,869,363,238 | 0.06 | 679 | **95.18** |
| | opt1 | $M_1$ | 1,088,946 | | | 3,235,842,391 | 0.02 | 2,972 | **98.99** |
| | | $M_2$ | | | | 2,938,993,549 | 0.02 | 2,699 | **99.54** |
| par-mne | orig | $M_1$ | 14,532,736 | 2 | 50 | 14,927,826,366 | 0.10 | 1,027 | **88.27** |
| | | $M_2$ | | | | 11,065,382,118 | 0.12 | 761 | **95.83** |
| | opt1 | $M_1$ | 1,088,944 | 2 | 50 | 3,583,085,865 | 0.02 | 3,290 | **99.68** |
| | | $M_2$ | | | | 3,255,485,394 | 0.01 | 2,989 | **99.66** |
| | opt2 | $M_1$ | 1,074,785 | 14,161 | 212,435 | 4,218,389,373 | 0.03 | 3,241 | **99.42** |
| | | $M_2$ | | | | 4,471,109,416 | 0.02 | 3,436 | **94.98** |
| scr | orig | $M_1$ | 4,350,760 | 1 | 31 | 6,445,643,039 | 0.03 | 1,481 | **89.97** |
| | | $M_2$ | | | 127 | 9,359,769,396 | 0.01 | 2,151 | **97.13** |
| JEDI | | $M_1$ | | | | | | | AVG: **91.55** |
| | | $M_2$ | | | | | | | AVG: **90.34** |
| OpenJDK | | $M_1$ | | | | | | | AVG: **94.82** |
| | | $M_2$ | | | | | | | AVG: **96.73** |

parallel streams, we attach PAPI to both the main thread and the workers involved in stream processing. We compute the baseline as the total cycles elapsed by all threads involved in stream processing.

While measuring the baseline, each cycle reading incurs a cost that is (partially) included in the measurement. However, this cost is in the order of the inner cost (i.e., $10^2$ cycles), small enough to be ignored when compared to the total cycles measured for a baseline.

#### 4.2.3 Accuracy Computation

Table 3 summarizes the results of our accuracy and profiling overhead evaluation in both the original (*orig*) and optimized versions (*opt*1 and *opt*2), discriminated in column *Ver.*, of mnemonics (mne), par-mnemonics (par-mne), and scrabble (scr) on $M_1$ and $M_2$. For each workload, we first report the number of anonymous (*#Anon. spans*), primordial (*#Prim. spans*), and support (*#Supp. spans*) spans (the latter two only for workloads executing parallel streams). We report the total elapsed cycles (*Total cycles*) and the corresponding CV, both computed from averages of 100 baseline measurements. We also report the CPS for each workload, which is computed as the total cycles divided by the total number of spans (rounded to an integer).

To estimate accuracy, we compute the *relative error (RE)*, considering the average baseline and average compensated cycles of 100 runs, such that accuracy is the quantity 1 – RE (shown as percentage). A RE equal to 0 would indicate the highest possible accuracy (100%).

We also report the average accuracy achieved for both the workloads from JEDI and OpenJDK at the bottom of the table. Due to space constraints, we report the complete evaluation results for both groups of workloads in Appendix D.

The evaluation results show that the average accuracy achieved on $M_1$ (92.72%) and $M_2$ (92.97%) are comparable, a sign of the portability of our approach. We achieve high accuracy for most of the studied workloads.

**Profiling and Optimizing Java Streams**

In Renaissance the lowest accuracies are obtained for the original mnemonics (i.e., 87.99% on $M_1$ and 95.18% on $M_2$), par-mnemonics (i.e., 88.27% on $M_1$ and 95.83% on $M_2$), and scrabble (i.e., 89.97% on $M_1$ and 97.13% on $M_2$). This drop in accuracy can be explained by the fact that these workloads exhibit a low CPS, such that the relative cost paid for the execution of the instrumentation code is more noticeable than for workloads with a higher CPS. Indeed, we find that there is a positive correlation between CPS and accuracy (*Pearson correlation coefficient* PCC: 0.71, considering all Renaissance workloads on the two machines). We remark that the correlation of CPS with accuracy (and with overhead, see Sec. 4.3) only holds for workloads exclusively performing stream processing. We also note that the CV for the total cycles elapsed in both mnemonics (i.e., 0.13 on $M_1$ and 0.06 on $M_2$) and par-mnemonics (i.e., 0.10 on $M_1$ and 0.12 on $M_2$) indicate an intrinsic variability in these workloads to which our evaluation is subjected.

We highlight that compensating for the instrumentation costs significantly helps increasing profile accuracy. For instance, we measured that the original profiles (without any compensation) show an accuracy as low as 49.11% for mnemonics, 53.61% for par-mnemonics, and 65.94% for scrabble. As shown in the table, our compensation technique notably increases the profile accuracy. The lack of compensation may result in misleading profiles, incorrectly pinpointing streams whose optimization may not have much impact on application performance, emphasizing the importance of a carefully tuned instrumentation and the use of perturbation-compensation techniques.

Our approach also achieves high average accuracy in both the workloads from JEDI (91.55% on $M_1$ and 90.34% on $M_2$) and OpenJDK (94.82% on $M_1$ and 96.73% on $M_2$).

We do not observe any significant variation in accuracy between workloads executing sequential streams and those additionally executing parallel streams, a sign that the execution mode has no major impact on accuracy.

As shown in Table 3, scrabble executes more cycles than the version of par-mnemonics incorporating our two optimizations, but takes less time to complete. This is explained by the fact that scrabble achieves almost a perfect load balance (CV of 0.03 on $M_1$ and 0.04 on $M_2$). Moreover, we find that workers often block (due to task synchronization) in the aforementioned optimized version of par-mnemonics both on $M_1$ and $M_2$, which is not the case for scrabble.

Overall, our evaluation shows an average accuracy of 92.85% across all evaluated workloads on $M_1$ and $M_2$, showing that StreamProf is suitable to perform cycle-accurate stream profiling. We discuss limitations of our compensation technique in Sec. 5.

### 4.3 Profiling Overhead

Here, we discuss the profiling overhead of StreamProf, which is reported in Table 4.

The overhead is computed as the *slowdown factor*, i.e., the execution time of the workload (which is reported in the table in column *Time*) with profiling divided by the execution time of the workload without any profiling. The reported values are the average of 100 runs. We also include the 95% CI.

In general, the highest overhead is mainly experienced while profiling the original mnemonics (i.e., 1.55× on $M_1$ and 1.46× on $M_2$), par-mnemonics (i.e., 1.51× on $M_1$

■ **Table 4** Results of the overhead evaluation.

| Bench. | Ver. | Machine | Time [ms] | Overhead Factor | 95% CI |
|---|---|---|---|---|---|
| mne | orig | $M_1$ | 4,994.72 | **1.55** | (1.51, 1.60) |
|  |  | $M_2$ | 3,303.64 | **1.46** | (1.44, 1.49) |
|  | opt1 | $M_1$ | 1,200.70 | **1.16** | (1.15, 1.16) |
|  |  | $M_2$ | 981.97 | **1.12** | (1.12, 1.13) |
| par-mne | orig | $M_1$ | 4,419.53 | **1.51** | (1.47, 1.54) |
|  |  | $M_2$ | 2,977.83 | **1.41** | (1.38, 1.45) |
|  | opt1 | $M_1$ | 1,106.22 | **1.13** | (1.13, 1.14) |
|  |  | $M_2$ | 905.19 | **1.11** | (1.11, 1.12) |
|  | opt2 | $M_1$ | 880.09 | **1.10** | (1.10, 1.11) |
|  |  | $M_2$ | 764.63 | **1.09** | (1.08, 1.09) |
| scr | orig | $M_1$ | 310.84 | **1.43** | (1.42, 1.44) |
|  |  | $M_2$ | 187.09 | **1.23** | (1.22, 1.23) |
| JEDI |  | $M_1$ |  | AVG: **1.13** |  |
|  |  | $M_2$ |  | AVG: **1.15** |  |
| OpenJDK |  | $M_1$ |  | AVG: **1.07** |  |
|  |  | $M_2$ |  | AVG: **1.06** |  |

and 1.41× on $M_2$), and scrabble (i.e., 1.43× on $M_1$ and 1.23× on $M_2$). This overhead is explained because these workloads have low CPS. Indeed, we find a negative correlation between CPS and overhead (PCC: −0.96, considering all Renaissance workloads on the two machines).

Our approach achieves low average overhead in both the workloads from JEDI (1.13× on $M_1$ and 1.15× on $M_2$) and OpenJDK (1.07× on $M_1$ and 1.06× on $M_2$).

Also, we find that there is no significant variation in the overhead between workloads executing sequential streams and the ones additionally executing parallel streams, a sign that the execution mode has little impact on the overhead.

Overall, our evaluation results show that StreamProf has an average overhead factor of 1.13×, across all evaluated workloads on $M_1$ and $M_2$. This profiling overhead enables the practical use of our tool (as shown in Sec. 3). We remark that our evaluation uses workloads that only execute streams, which can be considered a worst case scenario. For applications not only executing streams (likely the common case in Java applications), the profiling overhead is expected to be lower (i.e., the overhead would be negligible for phases of an application that do not execute streams).

## 5 Discussion

Here, we discuss limitations of our approach and threats to validity in our evaluation.

**Limitations of our technique.** While we estimate and compensate for some costs, the inserted instrumentation code may affect, e.g., Just-In-Time (JIT) compiler optimizations, register pressure, cache locality, and thread scheduling. These additional sources of perturbation are in practice very challenging to estimate and compensate. We also do not compensate for the extra cycles to maintain location information, be-cause its impact on accuracy and overhead is negligible in our measurements (below 1%). Nevertheless, our work introduces a perturbation-compensation technique to improve the accuracy of cycle profiling on the JVM.

The profiles produced using our technique are platform-dependent. A change in the underlying hardware (e.g., the use of a different processor) or software (e.g., the use of a different JCL), requires re-running our analysis to generate profiles valid in the new setting. We also note that our approach depends on the availability of per-thread virtualized reference-cycle counters.

We derived our instrumentation logic considering the implementation of the Stream API in OpenJDK [46]. Because of the design of the API, our instrumentation needs to target private classes within the JCL. Since private classes depend on a given implementation, our instrumentation may require changes to work on a different JCL, although we could easily adapt it. Furthermore, we have validated that our instrumentation is correct from Java version 8 (when the Stream API was introduced) up to version 17 (the latest long-term-support release at the time of writing).

Our definition of stream considers subtypes of `AbstractPipeline`, which implements the top-level interface of the Stream API, i,e., java.util.stream.BaseStream. These classes provide an entry and exit point for all stream execution methods, which are identified and profiled by `StreamProf`. We exclude entities that only implement the interface `BaseStream` without extending `AbstractPipeline` because `BaseStream` does not provide such entry points. As a result, it would be impossible to locate stream execution methods without knowledge provided by the user. Without information of the stream execution methods, it would not be possible for `StreamProf` to locate problematic stream executions. For completeness, as future work we plan to include ways to profile all stream execution methods of implementers of `BaseStream` that do not extend `AbstractPipeline`.

Our technique only tracks the stream-execution cycles, disregarding thread-scheduling overheads in the OS. As a result, we currently do not measure the number of thread context switches occurring during the execution of a span. A span incurring a high number of context switches (in proportion to the span's cycles) may indicate inefficient resource usage.

Our technique can guarantee the generation of complete stream profiles provided that the target application terminates. When profiling a continuously running application, only partial stream profiles can be reconstructed from the spans that ended before each data dumping.

**Threats to validity in the evaluation.** Our accuracy evaluation focuses on the total compensated cycles of all profiled spans (see Sec. 4), thus the accuracy of individual spans remains unknown. Unfortunately, evaluating span-level accuracy is challenging and may require dedicated hardware.

The results achieved by our approach may not be valid on other platforms. We conduct our experiments on two different machines (see Sec. 3 and Sec. 4) to confirm that the detected issues (and the speedups achieved) hold in different settings, indicating that the performance problems reported in this paper are not present on a single platform, but are related to the studied workloads.

Cycle measurements are subjected to variability. To mitigate this issue, we consider 100 runs for all the evaluations to report averages providing more reliable estimations.

## 6 Related Work

Here, we first discuss studies related to the use and optimization of Java streams and then we review work analyzing several aspects of an application running on the JVM.

**Use and optimization of Java streams.** Some authors have studied the use of streams in the wild. Tanaka et al. [47] mine 100 software repositories to study the use of lambda expressions, streams, and the java.util.Optional class. Khatchadourian et al. [5] examine 34 Java projects to study the use of streams in Java. This work represents the first attempt to specifically discover use cases and misuses of the Stream API. Nostas et al. [6] study the use of streams in 610 projects. This work is a partial replication of the study of Khatchadourian et al. considering a larger number of projects and validating the results previously obtained. Rosales et al. [48] conduct a large-scale study considering 132,211 open-source projects using the Stream API, mostly confirming the findings of the aforementioned work.

Prior work [49] argues the applicability of dynamic analysis to profile Java streams. To the best of our knowledge, only six studies have focused on the optimization of streams. Ishizaki et al. [14] modify the IBM J9 JVM [50] and the Testarossa compiler [51] to translate stream code into optimized GPU code. Hayashi et al. [15] extend this work, proposing a supervised machine-learning approach that allows the runtime to select between CPU or GPU to speed up stream execution. Differently from our work, which requires manual code inspection (and knowledge of the Stream API), the above approaches enable the automatic optimization of parallel streams mainly via compiler- and hardware-related support. However, the applicability of our technique is not limited to a proprietary compiler, but enables the optimization of streams in applications running on popular JVMs (e.g., HotSpot [52] and GraalVM [53]). Moreover, both approaches can optimize only integer-based parallel streams executed via the terminal operation forEach [54]. In contrast, StreamProf helps developers optimize parallel and sequential streams, either primitive-type or object-based, executed via any terminal operation. Differently from both approaches, we can also speed up stream processing without any special CPU/GPU support.

Khatchadourian et al. [10, 11] introduce an Eclipse plugin helping developers better code streams. As remarked by the authors, their approach is unable to assess the optimization of a stream considering input size/overhead trade-offs [11]. The plugin mainly relies on static analysis techniques that may not fully capture the runtime behavior of the analyzed streams. In contrast, our technique is the first that measures the computations performed by a stream. We consider the work of Khatchadourian et al. complementary to our work, as the plugin can automatically suggest safe pipeline refactoring to better leverage streams.

Møller et al. [9] use bytecode-level transformations to remove abstraction overheads of sequential streams. However, the approach cannot analyze parallel streams. Basso et al. [13] exploit bytecode-level transformations to specifically remove abstraction overheads of parallel streams. Unfortunately, both approaches overlook runtime information that can be crucial to spot stream-related performance issues. We consider the above approaches complementary to our work, as the optimized stream processing enabled by our profiles may further benefit from bytecode-level transformations.

**Analyses and profilers for the JVM.** Several authors study the dynamic behavior of applications on the JVM [17, 18, 20, 55–58]. While software debloating techniques [59, 60] for Java applications can remove unused fields, the optimization described in Sec. 3.2 targets a field and an argument that are actually used (via stream processing) within recursive code. Some approaches [61–63] focus on debugging big data systems (e.g., Apache Spark [64]) that similarly to the Stream API enable MapReduce-like [7] data transformations. While such work targets distributed systems, Java streams execute in a single JVM process. Lastly, there are many general-purpose profilers for the JVM [16, 19, 65–70]. None of the aforementioned studies or tools specifically addresses cycle-accurate profiling of Java streams.

## 7 Conclusions

In this paper, we presented a novel profiling technique for Java streams. Our technique measures the computations performed by sequential and parallel streams in terms of elapsed reference cycles, accurately capturing stream nesting, and producing actionable profiles that ease the detection and optimization of stream-related performance issues. StreamProf guides our optimizations in two stream-based workloads of the state-of-the-art benchmark suite Renaissance, enabling speedups up to a factor of 5×.

Our approach *compensates* for the extra cycles caused by the execution of inserted instrumentation code, i.e., we estimate those cycles and remove them to increase profile accuracy. Our evaluation results show that the accuracy of the profiles after compensation is 92.85% on average. Our compensation technique significantly improves accuracy in comparison to naive cycle profiling (e.g., for mnemonics, from an accuracy of 49.11% without compensation to an accuracy of 87.99% after compensation).

As part of our future work, we plan to apply our profiler to some open-source software projects and investigate stream-related optimization opportunities. We also plan to demonstrate the general applicability of our technique to third-party stream libraries targeting Java [71–73] and other programming languages [74–76]. We plan to measure per-span thread context switches to help unveil inefficient resource usage in stream executions. Also, we plan to support additional visualizations of stream profiles (e.g., pie charts showing the work distribution among the workers executing a parallel stream). We also plan to release StreamProf as open-source software.

**Data Availability Statement** The related artifact [77] consists of a Docker image that enables the production of the tables included in this paper (excluding appendixes).

**Acknowledgements** This work has been supported by Oracle (ERO project 1332) and by the Swiss National Science Foundation (project 200020_188688).

## A  Instrumentation Logic

Here, we detail the instrumentation logic used to detect stream executions as described in Sec. 2.1. In listing 2 we describe our approach using both data structures and python-based pseudocode that favors conciseness. We also use special variables as provided by DiSL [31], which we explain below.

**Listing 2**  Instrumentation logic in pseudocode.

```python
# LOCATION

# Variables used to maintain the method ID
@ThreadLocal int method_ID
@SyntheticLocal int method_ID_old

# Update the method ID upon method entry, excluding methods belonging to the
# Stream API and those that are guaranteed not to invoke any terminal operation
@Before(StreamExecutionMethod):
  method_ID_old = method_ID
  # Method new_method_ID() creates a new unique ID for a stream execution method
  method_ID = new_method_ID()

@After(StreamExecutionMethod):
  method_ID = method_ID_old

# SEQUENTIAL STREAMS - SPAN-BEGIN. Detect the execution of a sequential stream
@Before(SequentialStreamExecution):
  register_anonymous_span_begin(method_ID)

# PARALLEL STREAMS - SPAN-BEGIN. Detect the execution of a parallel stream
@Before(ParallelStreamExecution):
  # Method get_head() obtains a reference to the head of the pipeline
  head = get_head()

  # Method get_ID(..) obtains the stream ID associated with the head, if any.
  # Otherwise returns minus one
  stream_ID = get_ID(head)

  if stream_ID != -1:
    # The stream ID has been set previously (i.e., this is a support span)
    register_support_span_begin(method_ID, stream_ID)
  else:
    # This is the primordial span (i.e., this is the first task executing computations
    # on behalf of the parallel stream)

    # Method set_ID(..) associates a unique stream ID with the head and returns it
    stream_ID = set_ID(head)
    register_primordial_span_begin(method_ID, stream_ID)

# ANY STREAM - SPAN-END. Detect the end of the execution of any stream
@After(AnyStreamExecution):
  register_span_end()
```

**Profiling and Optimizing Java Streams****Thread-local variables.** Our instrumentation needs to store information related to the current thread. To this end, we leverage thread-local variables provided by DiSL, which allow efficiently storing information in an additional instance field of the object representing the current thread. In DiSL, a thread-local variable is declared with the `@ThreadLocal` annotation.

**Synthetic local variables.** Our instrumentation requires passing information between different code snippets woven into the same method body. To do so, we use synthetic local variables provided by DiSL, which are efficiently embedded within the frames of the Java call stack, avoiding heap allocations. In DiSL, a synthetic local variable is declared using the `@SyntheticLocal` annotation.

**Snippets.** We express our instrumentation using snippets as supported by DiSL. Note that each snippet is preceded by an annotation (e.g., `@Before`) and a predicate (e.g., `StreamExecutionMethod`), which in conjunction specify where the instrumentation code should be inserted. For instance, the aforementioned combination would insert the instrumentation code at the beginning of every stream execution method.

## B  Span Reconstruction Algorithm

Here, we detail the span-reconstruction algorithm used in the offline post-processing as described in Sec. 2.1. To ease its comprehension, in listing 3 we describe our approach using python-based pseudocode for conciseness.

**Listing 3** Span reconstruction algorithm in pseudocode.

```
 1  # DATA TYPES:
 2
 3  # Represents either a span-begin or span-end event as collected by the profiler
 4  Event:
 5    # Possible values:
 6    # ASB: anonymous span begin
 7    # SSB: support span begin
 8    # PSB: primordial span begin
 9    # SE: span end
10    Enum types
11    long cycles
12    int stream_ID
13    int method_ID
14
15  # Represents a reconstructed span from its respective span-begin and span-end events
16  Span:
17    long cycles_begin
18    long cycles_end
19    int stream_ID
20    int method_ID
21    int thread_ID
22    boolean is_primordial = False
23    long nested_cycles
24    int nested_anonymous_spans
25    int nested_support_spans
```

10:24

```
26    int nested_primordial_spans
27    Span parent = None # valid only within the same thread
28    Span primordial = None # the primordial span of this span, if any
29    int nesting_level = -1
30
31    # Constructor
32    Span(cycles_begin, stream_ID, method_ID, thread_ID, is_primordial):
33       # Initializations omitted
34
35    # Return the measured cycles of this span
36    measured_cycles():
37       return (cycles_end - cycles_begin)
38
39    # Return the compensated cycles of this span
40    # OC_ANON,OC_PRIM, OC_SUPP and IC are the costs as estimated in the calibration phase
41    compensated_cycles():
42       return (measured_cycles() - nested_cycles)
43          - (nested_anonymous_spans * OC_ANON)
44          - (nested_primordial_spans * OC_PRIM)
45          - (nested_support_spans * OC_SUPP)
46          - IC
47
48    # Return the enclosing span (None for non-nested streams)
49    outer_span():
50       if primordial:
51          return primordial.parent
52       else:
53          return parent
54
55 # Represents the profile of a thread involved in stream processing
56 ThreadProfile:
57    int thread_ID
58    str thread_name
59    # List of all events in the trace corresponding to this thread profile
60    events = [Event]
61    # Auxiliary stack to process the events corresponding to this thread profile
62    Stack stack
63    # List of all reconstructed spans executed by this thread profile
64    spans = [Span]
65    # Maps a stream ID to the list of the stream's named spans in this thread profile
66    named_spans = dict{int, [Span]}
67
68    # Constructor
69    Thread(thread_ID, thread_name, events):
70       # Initializations omitted
71       compute_thread_spans()
72
73 # Convert the list of events of this thread profile to a list of reconstructed spans
74    compute_thread_spans():
75       for event in events:
76          match event.types:
77             case ASB: # Anonymous span begin
78                stack.push(Span(event.cycles, -1, event.method_ID, thread_ID, False))
```

```
79              case SSB: # Support span begin
80                stack.push(Span(event.cycles, event.streamID, event.method_ID, thread_ID, False))
81              case PSB: # Primordial span begin
82                stack.push(Span(event.cycles, event.streamID, event.method_ID, thread_ID, True))
83              case SE: # Any span end
84                span = stack.pop()
85                span.end_cycles = event.cycles
86                if not stack.is_empty(): # The span is nested
87                  parent_span = stack.peek()
88                  span.parent = parent_span
89                  parent_span.nested_cycles += span.measured_cycles()
90                  if span.stream_ID == -1:
91                    parent_span.nested_anonymous_spans += 1
92                  else:
93                    named_spans[stream_ID].append(span)
94                    if span.is_primordial:
95                      parent_span.nested_primordial_spans += 1
96                    else:
97                      parent_span.nested_support_spans += 1
98                spans.append(span)
99
100 # The profile of an application considering all its corresponding thread profiles
101 ApplicationProfile:
102   # Initialization omitted (contains all spans in all thread profiles)
103   all_spans = [Span]
104
105   # Maps a stream ID to the list of the stream's named spans across all thread profiles
106   merged_named_spans = dict{int, [Span]}
107
108   # Constructor
109   ApplicationProfile(thread_profiles):
110     merge_named_spans(thread_profiles)
111     update_primordial()
112     update_nesting_levels()
113
114   # Merge the named spans of a parallel stream across all thread profiles
115   merge_named_spans(thread_profiles):
116     for tp in thread_profiles:
117       for k, v in tp.named_spans:
118         merged_named_spans[k].append[v]
119
120   # Update the primordial span in all support spans of the same parallel stream
121   update_primordial():
122     for spans in merged_named_spans.values():
123       # Method findPrimordialSpan(..) returns the unique primordial span
124       primordial = findPrimordialSpan(spans)
125       for span in spans:
126         if not span.is_primordial:
127           span.primordial = primordial
128
129   # Compute the nesting level of a span. If the span is nested, recursively computes the
130   # chain of outer spans until the outermost span is found. Otherwise, assigns zero
131   compute_nesting_level(span):
```

```
132     if span.nesting_level = -1:
133       if span.outer_span:
134         span.nesting_level = compute_nesting_level(span.outer_span()) + 1
135       else:
136         span.nesting_level = 0 # The span is not nested
137     return span.nesting_level
138
139   # Update the nesting levels of all spans after merging all thread profiles
140   update_nesting_levels():
141     for span in all_spans:
142       compute_nesting_level(span)
143
144   # Return the stream-related compensated cycles of this application profile
145   get_compensated_cycles():
146     return sum(span.compensated_cycles() for span in all_spans)
147
148 # SAMPLE USE:
149
150 # An example reporting the total compensated stream-processing cycles of an application
151 main():
152   thread_profiles = [ThreadProfile]
153
154   # For each trace dumped by the profiler, method retrieveTraces() retrieves the
155   # corresponding sequence of span-begin and span-end events in the trace along with the
156   # information of the associated thread. The definition of the data type "trace" is omitted.
157   for trace in retrieveTraces():
158     for thread_name, thread_ID, events in trace:
159       thread_profiles.append(ThreadProfile(thread_name, thread_ID, events))
160
161   application_profile = ApplicationProfile(thread_profiles)
162
163   print(application_profile.get_compensated_cycles())
```

## C  Heatmaps

Here, we show the customized heatmaps reported by StreamProf to guide the optimizations described in Sec. 3.

Fig. 3 shows two customized heatmaps produced by StreamProf from the profiles generated for mnemonics.[14] We use customized heatmaps to condense the information profiled (e.g., stream-nesting information, compensated cycles, the number of stream executions, the presence of hot regions), helping developers analyze stream processing.

The first heatmap (Fig. 3a) corresponds to the execution of the original (unmodified) version. The heatmap divides streams into groups. On the x-axis, streams are grouped by their (compensated) cycles as reported by StreamProf. On the y-axis, streams are grouped by their nesting level. mnemonics accounts for 44 nesting levels, which we aggregate in groups of 10 (starting from 0), to ease visualization. The number reported in each cell is the number of executed streams for a given range of nesting levels and cycles. Differently from a traditional heatmap, the color of a cell indicates the total cycles of all the streams in the group. The darker a cell, the higher the total cycles elapsed by the streams in the group.

---

[14] We only show the profiles obtained for mnemonics on $M_1$, which are very similar to those obtained on $M_2$.

As shown in Fig. 3a, there is a substantial amount of stream executions at various nesting levels having a major contribution to the cycles profiled in mnemonics. In our approach, we focus on the darker regions indicated by the heatmap, as optimizations that reduce the cycles in these regions are likely to improve performance. To this end, StreamProf complementary reports the locations of the streams responsible for most of the stream processing (as shown in Sec. 3).

Fig. 3b shows the heatmap of the workload after our optimizations. A visual comparison between Fig. 3a and Fig. 3b highlights how the removal of problematic streams in the two analyzed hot locations avoids unneeded and expensive computations in the optimized version.

We note that the small differences in the number of stream executions per category reported in Fig. 3a and Fig. 3b are explained by a variability in the benchmark executions to which our experiments are subjected (as discussed in Sec. 5). As a result, in different benchmark runs, a few stream executions may be reported in different (typically neighboring) categories.

## D  Additional Evaluation Results

Here, we show the complete results of our accuracy and overhead evaluation on both the workloads from JEDI and OpenJDK, which are discussed in Sec. 4. We report averages at the bottom of the all the tables presented in this appendix.

We note that most of the workloads from OpenJDK execute in a short time as they were designed to exercise only specific features of the Stream API.

Due to space constraints, we use short names to identify each workload from OpenJDK in Table 7 and Table 8. The original names used in the OpenJDK18 open-source repository [44] are as follows: w1: AnyMatcher, w2: AllMatcher, w3: Decomposition.saturated, w4: IntegerSum, w5: PrimesSieve, w6: IntegerMax, w7: IntegerDuplicate, w8: PrimesFilter.t100, w9-par: Decomposition.thinktime_par, w10-par: IntegerSum_par, w11-par: IntegerMax_par, w12-par: IntegerDuplicate_par, w13-par: PrimesFilter.t10000_par.

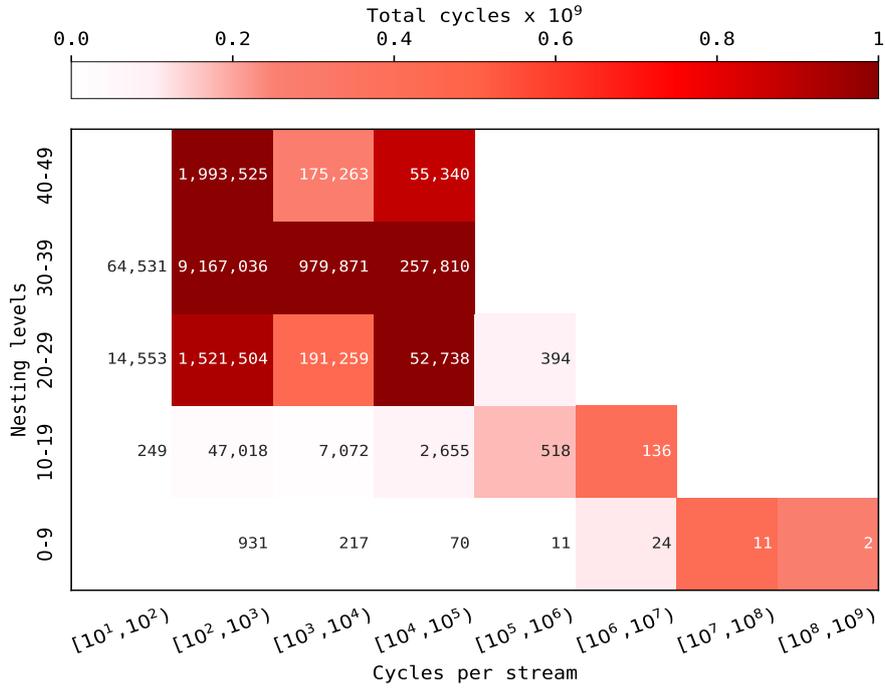

**(a)** Original version

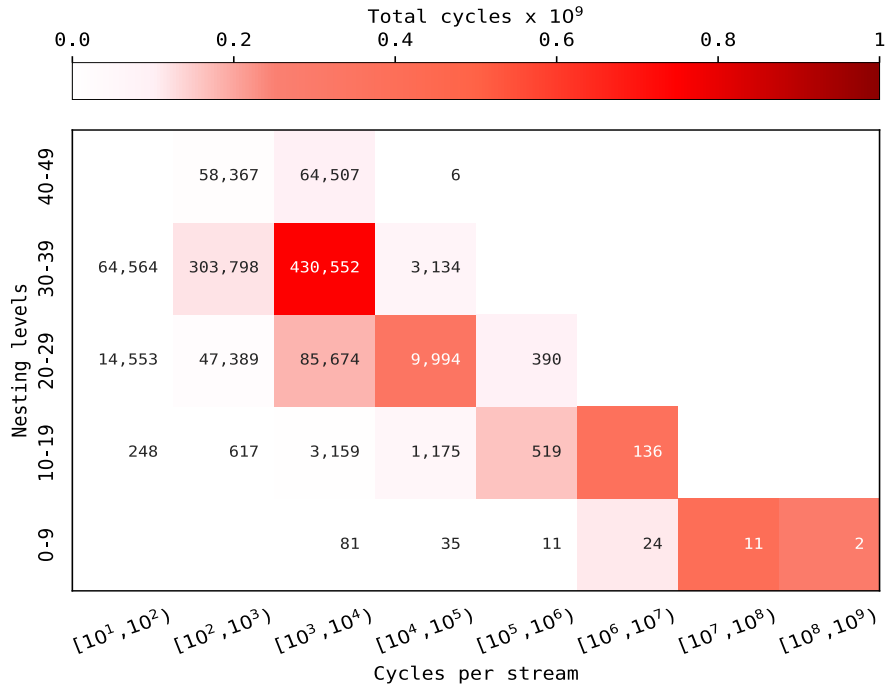

**(b)** Optimized version

**Figure 3** Stream execution before and after optimizations in mnemonics. Each cell shows the number of executed streams in the corresponding categories, whereas the cell color corresponds to the total cycles of these streams.

**Table 5** Results of the accuracy evaluation on the workloads from JEDI.

| Bench. | Machine | #Anon. Spans | #Prim. Spans | #Supp. Spans | Total cycles AVG | CV | CPS | Accuracy [%] |
|---|---|---|---|---|---|---|---|---|
| q1 | $M_1$ | 2 | | | 1,721,311,380 | 0.01 | 860,655,690 | **92.82** |
| | $M_2$ | 2 | | | 2,383,530,808 | 0.09 | 1,191,765,404 | **93.31** |
| q1-par | $M_1$ | 1 | 1 | 31 | 2,394,566,469 | 0.01 | 72,562,620 | **90.13** |
| | $M_2$ | 1 | 1 | 62 | 2,809,949,167 | 0.09 | 43,905,456 | **91.63** |
| q2 | $M_1$ | 10 | | | 370,840,284 | 0.02 | 37,084,028 | **92.93** |
| | $M_2$ | 10 | | | 491,430,575 | 0.07 | 49,143,058 | **98.01** |
| q3 | $M_1$ | 4 | | | 1,801,202,558 | 0.10 | 450,300,640 | **91.18** |
| | $M_2$ | 4 | | | 2,660,667,317 | 0.10 | 665,166,829 | **92.15** |
| q3-par | $M_1$ | 1 | 3 | 93 | 2,607,513,919 | 0.04 | 26,881,587 | **88.51** |
| | $M_2$ | 1 | 3 | 140 | 2,981,862,225 | 0.06 | 20,707,377 | **87.41** |
| q4 | $M_1$ | 3 | | | 1,235,565,394 | 0.03 | 411,855,131 | **89.36** |
| | $M_2$ | 3 | | | 1,638,539,192 | 0.09 | 546,179,731 | **83.71** |
| q5 | $M_1$ | 7 | | | 1,936,425,328 | 0.01 | 276,632,190 | **95.27** |
| | $M_2$ | 7 | | | 2,054,445,608 | 0.03 | 293,492,230 | **81.71** |
| q6 | $M_1$ | 1 | | | 567,249,098 | 0.01 | 567,249,098 | **90.30** |
| | $M_2$ | 1 | | | 870,871,942 | 0.08 | 870,871,942 | **90.45** |
| q7 | $M_1$ | 31 | | | 1,865,036,543 | 0.01 | 60,162,469 | **82.09** |
| | $M_2$ | 31 | | | 2,476,180,358 | 0.10 | 79,876,786 | **84.36** |
| q8 | $M_1$ | 9 | | | 2,161,629,272 | 0.02 | 240,181,030 | **96.90** |
| | $M_2$ | 9 | | | 2,439,844,650 | 0.03 | 271,093,850 | **96.89** |
| q9 | $M_1$ | 7 | | | 3,229,724,299 | 0.06 | 461,389,186 | **97.66** |
| | $M_2$ | 7 | | | 3,574,183,258 | 0.04 | 510,597,608 | **91.36** |
| q10 | $M_1$ | 1,685,944 | | | 2,513,228,294 | 0.01 | 1,491 | **83.84** |
| | $M_2$ | 1,685,944 | | | 3,156,714,042 | 0.05 | 1,872 | **84.67** |
| q11 | $M_1$ | 7 | | | 272,898,418 | 0.01 | 38,985,488 | **99.49** |
| | $M_2$ | 7 | | | 401,778,975 | 0.08 | 57,396,996 | **99.35** |
| q12 | $M_1$ | 3 | | | 1,249,998,343 | 0.01 | 416,666,114 | **89.97** |
| | $M_2$ | 3 | | | 1,861,945,683 | 0.11 | 620,648,561 | **92.91** |
| q13 | $M_1$ | 4 | | | 2,731,782,055 | 0.02 | 682,945,514 | **92.42** |
| | $M_2$ | 4 | | | 3,001,892,492 | 0.01 | 750,473,123 | **85.93** |
| q14 | $M_1$ | 200,002 | | | 1,147,485,490 | 0.03 | 5,737 | **87.10** |
| | $M_2$ | 200,002 | | | 1,434,930,717 | 0.09 | 7,175 | **86.87** |
| q15 | $M_1$ | 6 | | | 759,864,247 | 0.10 | 126,644,041 | **84.65** |
| | $M_2$ | 6 | | | 1,115,776,450 | 0.14 | 185,962,742 | **92.10** |
| q16 | $M_1$ | 4 | | | 315,863,876 | 0.04 | 78,965,969 | **98.16** |
| | $M_2$ | 4 | | | 387,515,417 | 0.05 | 96,878,854 | **99.09** |
| q17 | $M_1$ | 4 | | | 4,917,510,274 | 0.01 | 1,229,377,568 | **98.59** |
| | $M_2$ | 4 | | | 5,356,269,858 | 0.02 | 1,339,067,465 | **95.39** |
| q18 | $M_1$ | 6 | | | 3,521,113,897 | 0.02 | 586,852,316 | **95.21** |
| | $M_2$ | 6 | | | 6,800,786,817 | 0.02 | 1,133,464,469 | **87.52** |
| q18-par | $M_1$ | 2 | 4 | 124 | 5,740,561,404 | 0.01 | 44,158,165 | **95.16** |
| | $M_2$ | 2 | 4 | 202 | 6,800,786,817 | 0.02 | 32,696,090 | **97.52** |
| q19 | $M_1$ | 214,379 | | | 1,505,820,737 | 0.08 | 7,024 | **88.35** |
| | $M_2$ | 214,379 | | | 1,829,283,858 | 0.06 | 8,533 | **86.41** |
| q20 | $M_1$ | 553,216 | | | 2,387,540,569 | 0.01 | 4,316 | **88.78** |
| | $M_2$ | 553,216 | | | 3,121,469,875 | 0.06 | 5,642 | **85.40** |
| q21 | $M_1$ | 322,633 | | | 5,031,578,786 | 0.05 | 15,595 | **85.30** |
| | $M_2$ | 322,633 | | | 6,503,770,125 | 0.06 | 20,158 | **82.26** |
| q22 | $M_1$ | 5 | | | 345,374,176 | 0.03 | 69,074,835 | **94.49** |
| | $M_2$ | 5 | | | 476,944,467 | 0.05 | 95,388,893 | **92.09** |
| Average | $M_1$ | | | | | | | AVG: **91.55** |
| | $M_2$ | | | | | | | AVG: **90.34** |

**Table 6** Results of the overhead evaluation on the workloads from JEDI.

| Bench. | Machine | Time [ms] | Overhead Factor | 95% CI |
|---|---|---|---|---|
| q1 | $M_1$ | 639.71 | **1.06** | (1.05, 1.08) |
|  | $M_2$ | 795.27 | **1.06** | (1.05, 1.07) |
| q1-par | $M_1$ | 113.82 | **1.07** | (1.06, 1.09) |
|  | $M_2$ | 115.71 | **1.06** | (1.05, 1.07) |
| q2 | $M_1$ | 137.70 | **1.07** | (1.06, 1.08) |
|  | $M_2$ | 163.99 | **1.02** | (1.01, 1.03) |
| q3 | $M_1$ | 669.49 | **1.18** | (1.17, 1.19) |
|  | $M_2$ | 887.74 | **1.16** | (1.15, 1.17) |
| q3-par | $M_1$ | 154.05 | **1.21** | (1.20, 1.22) |
|  | $M_2$ | 146.27 | **1.22** | (1.21, 1.23) |
| q4 | $M_1$ | 459.69 | **1.23** | (1.22, 1.25) |
|  | $M_2$ | 546.71 | **1.30** | (1.29, 1.31) |
| q5 | $M_1$ | 718.53 | **1.03** | (1.01, 1.04) |
|  | $M_2$ | 685.46 | **1.22** | (1.21, 1.23) |
| q6 | $M_1$ | 211.11 | **1.09** | (1.07, 1.10) |
|  | $M_2$ | 290.61 | **1.10** | (1.09, 1.11) |
| q7 | $M_1$ | 693.14 | **1.28** | (1.27, 1.30) |
|  | $M_2$ | 826.27 | **1.33** | (1.32, 1.34) |
| q8 | $M_1$ | 802.20 | **1.01** | (1.01, 1.02) |
|  | $M_2$ | 814.05 | **1.02** | (1.01, 1.03) |
| q9 | $M_1$ | 1198.78 | **1.02** | (1.01, 1.03) |
|  | $M_2$ | 1192.47 | **1.09** | (1.08, 1.11) |
| q10 | $M_1$ | 933.67 | **1.35** | (1.34, 1.36) |
|  | $M_2$ | 1053.26 | **1.35** | (1.34, 1.36) |
| q11 | $M_1$ | 101.35 | **1.04** | (1.03, 1.05) |
|  | $M_2$ | 134.15 | **1.01** | (1.01, 1.03) |
| q12 | $M_1$ | 465.01 | **1.13** | (1.12, 1.14) |
|  | $M_2$ | 621.46 | **1.07** | (1.06, 1.08) |
| q13 | $M_1$ | 1013.51 | **1.08** | (1.07, 1.09) |
|  | $M_2$ | 1001.57 | **1.16** | (1.15, 1.17) |
| q14 | $M_1$ | 426.88 | **1.29** | (1.28, 1.32) |
|  | $M_2$ | 478.98 | **1.30** | (1.29, 1.31) |
| q15 | $M_1$ | 282.71 | **1.15** | (1.14, 1.16) |
|  | $M_2$ | 372.36 | **1.10** | (1.09, 1.11) |
| q16 | $M_1$ | 117.30 | **1.05** | (1.04, 1.06) |
|  | $M_2$ | 129.84 | **1.02** | (1.01, 1.03) |
| q17 | $M_1$ | 1824.83 | **1.03** | (1.02, 1.04) |
|  | $M_2$ | 1787.06 | **1.14** | (1.13, 1.15) |
| q18 | $M_1$ | 1307.56 | **1.05** | (1.04, 1.06) |
|  | $M_2$ | 496.35 | **1.15** | (1.14, 1.17) |
| q18-par | $M_1$ | 480.67 | **1.07** | (1.06, 1.08) |
|  | $M_2$ | 496.35 | **1.05** | (1.04, 1.06) |
| q19 | $M_1$ | 559.79 | **1.21** | (1.20, 1.23) |
|  | $M_2$ | 610.61 | **1.21** | (1.20, 1.22) |
| q20 | $M_1$ | 886.50 | **1.24** | (1.23, 1.25) |
|  | $M_2$ | 1041.46 | **1.25** | (1.24, 1.27) |
| q21 | $M_1$ | 1869.16 | **1.24** | (1.23, 1.25) |
|  | $M_2$ | 2170.20 | **1.25** | (1.24, 1.26) |
| q22 | $M_1$ | 128.28 | **1.07** | (1.05, 1.09) |
|  | $M_2$ | 159.24 | **1.11** | (1.10, 1.12) |
| Average | $M_1$ |  | AVG: **1.13** |  |
|  | $M_2$ |  | AVG: **1.15** |  |

**Table 7** Results of the accuracy evaluation on the workloads from OpenJDK.

| Bench. | Machine | #Anon. Spans | #Prim. Spans | #Supp. Spans | Total cycles AVG | CV | CPS | Accuracy [%] |
|---|---|---|---|---|---|---|---|---|
| w1 | $M_1$ | 1 | | | 879,573 | 0.04 | 879,573 | **93.15** |
|    | $M_2$ | 1 | | | 753,280 | 0.05 | 753,280 | **95.05** |
| w2 | $M_1$ | 1 | | | 1,242,508 | 0.01 | 1,242,508 | **94.40** |
|    | $M_2$ | 1 | | | 1,078,703 | 0.01 | 1,078,703 | **94.40** |
| w3 | $M_1$ | 1 | | | 7,240,921 | 0.01 | 7,240,921 | **98.68** |
|    | $M_2$ | 1 | | | 6,271,850 | 0.01 | 6,271,850 | **99.78** |
| w4 | $M_1$ | 1 | | | 135,089,803 | 0.01 | 135,089,803 | **98.39** |
|    | $M_2$ | 1 | | | 152,494,128 | 0.01 | 152,494,128 | **99.50** |
| w5 | $M_1$ | 1 | | | 160,450,498 | 0.10 | 160,450,498 | **99.21** |
|    | $M_2$ | 1 | | | 151,542,390 | 0.09 | 151,542,390 | **99.53** |
| w6 | $M_1$ | 1 | | | 216,617,925 | 0.01 | 216,617,925 | **91.78** |
|    | $M_2$ | 1 | | | 197,130,665 | 0.01 | 197,130,665 | **95.54** |
| w7 | $M_1$ | 10,485,761 | | | 2,491,458,725 | 0.13 | 238 | **94.20** |
|    | $M_2$ | 10,485,761 | | | 2,326,973,493 | 0.14 | 222 | **92.74** |
| w8 | $M_1$ | 1 | | | 3,038,972,058 | 0.01 | 3,038,972,058 | **99.89** |
|    | $M_2$ | 1 | | | 2,857,904,263 | 0.01 | 2,857,904,263 | **99.99** |
| w9-par | $M_1$ | | 1 | 31 | 9,752,180 | 0.14 | 304,756 | **89.63** |
|        | $M_2$ | | 1 | 38 | 9,280,780 | 0.12 | 237,969 | **88.07** |
| w10-par | $M_1$ | | 1 | 31 | 166,038,120 | 0.07 | 5,188,691 | **86.36** |
|         | $M_2$ | | 1 | 31 | 137,748,443 | 0.05 | 4,304,639 | **97.85** |
| w11-par | $M_1$ | | 1 | 31 | 202,450,863 | 0.04 | 6,326,589 | **93.29** |
|         | $M_2$ | | 1 | 31 | 131,721,768 | 0.03 | 4,116,305 | **98.44** |
| w12-par | $M_1$ | | 1 | 31 | 337,715,623 | 0.05 | 10,553,613 | **93.67** |
|         | $M_2$ | | 1 | 31 | 233,671,930 | 0.05 | 7,302,248 | **97.03** |
| w13-par | $M_1$ | | 1 | 35 | 3,041,238,569 | 0.01 | 84,478,849 | **99.99** |
|         | $M_2$ | | 1 | 35 | 2,850,698,723 | 0.01 | 79,186,076 | **99.52** |
| Average | $M_1$ | | | | | | | AVG: **94.82** |
|         | $M_2$ | | | | | | | AVG: **96.73** |

**Eduardo Rosales, Matteo Basso, Andrea Rosà, and Walter Binder**■ **Table 8** Results of the overhead evaluation on the workloads from OpenJDK.

| Bench. | Machine | Time [ms] | Overhead Factor | 95% CI |
|---|---|---|---|---|
| w1 | $M_1$ | 0.32 | **1.01** | (1.01, 1.02) |
|    | $M_2$ | 0.24 | **1.05** | (1.04, 1.06) |
| w2 | $M_1$ | 0.45 | **1.02** | (1.01, 1.03) |
|    | $M_2$ | 0.35 | **1.02** | (1.01, 1.04) |
| w3 | $M_1$ | 2.68 | **1.01** | (1.01, 1.03) |
|    | $M_2$ | 2.10 | **1.01** | (1.01, 1.02) |
| w4 | $M_1$ | 50.23 | **1.02** | (1.01, 1.03) |
|    | $M_2$ | 50.90 | **1.02** | (1.01, 1.03) |
| w5 | $M_1$ | 66.43 | **1.01** | (1.01, 1.02) |
|    | $M_2$ | 50.58 | **1.01** | (1.01, 1.02) |
| w6 | $M_1$ | 80.38 | **1.02** | (1.01, 1.04) |
|    | $M_2$ | 65.80 | **1.02** | (1.02, 1.03) |
| w7 | $M_1$ | 829.77 | **1.37** | (1.35, 1.38) |
|    | $M_2$ | 769.94 | **1.33** | (1.31, 1.34) |
| w8 | $M_1$ | 1126.32 | **1.01** | (1.01, 1.03) |
|    | $M_2$ | 953.05 | **1.01** | (1.01, 1.02) |
| w9-par | $M_1$ | 0.75 | **1.07** | (1.05, 1.08) |
|        | $M_2$ | 0.77 | **1.11** | (1.09, 1.12) |
| w10-par | $M_1$ | 8.82 | **1.35** | (1.33, 1.36) |
|         | $M_2$ | 6.64 | **1.22** | (1.21, 1.23) |
| w11-par | $M_1$ | 10.65 | **1.03** | (1.03, 1.04) |
|         | $M_2$ | 6.30 | **1.02** | (1.02, 1.03) |
| w12-par | $M_1$ | 17.87 | **1.03** | (1.02, 1.04) |
|         | $M_2$ | 11.02 | **1.01** | (1.01, 1.02) |
| w13-par | $M_1$ | 238.15 | **1.01** | (1.01, 1.03) |
|         | $M_2$ | 194.82 | **1.01** | (1.01, 1.02) |
| Average | $M_1$ |  | **1.07** |  |
|         | $M_2$ |  | **1.06** |  |

## References

[1] Oracle. *Package java.util.stream*. https://docs.oracle.com/en/java/javase/19/docs/api/java.base/java/util/stream/Stream.html. 2022. (Visited on 2022-12-30).

[2] Richard Bird and Philip Wadler. *An Introduction to Functional Programming*. 1st. GBR: Prentice Hall International (UK) Ltd., 1988. ISBN: 0134841891.

[3] Joshua Bloch. *Effective Java*. 3rd. Addison-Wesley, 2018. ISBN: 978-0-13-468599-1.

[4] Davood Mazinanian, Ameya Ketkar, Nikolaos Tsantalis, and Danny Dig. "Understanding the Use of Lambda Expressions in Java". In: *Proc. ACM Program. Lang.* 1.OOPSLA (Oct. 2017), pages 1–31. ISSN: 2475-1421. DOI: 10.1145/3133909.

[5] Raffi Khatchadourian, Yiming Tang, Mehdi Bagherzadeh, and Baishakhi Ray. "An Empirical Study on the Use and Misuse of Java 8 Streams". In: *FASE*. 2020, pages 97–118. DOI: 10.1007/978-3-030-45234-6_5.10:33

**Profiling and Optimizing Java Streams**[62]  Muhammad Ali Gulzar, Matteo Interlandi, Tyson Condie, and Miryung Kim. "Debugging Big Data Analytics in Spark with BigDebug". In: *SIGMOD*. 2017, 1627–1630. DOI: 10.1145/3035918.3058737.

[63]  Muhammad Ali Gulzar, Siman Wang, and Miryung Kim. "BigSift: Automated Debugging of Big Data Analytics in Data-Intensive Scalable Computing". In: *ESEC/FSE*. 2018, 863–866. DOI: 10.1145/3236024.3264586.

[64]  Matei Zaharia, Mosharaf Chowdhury, Tathagata Das, Ankur Dave, Justin Ma, Murphy McCauley, Michael Franklin, Scott Shenker, and Ion Stoica. "Resilient Distributed Datasets: A Fault-tolerant Abstraction for In-memory Cluster Computing". In: *NSDI'12: Proceedings of the 9th USENIX Conference on Networked Systems Design and Implementation*. USENIX Association, 2012, 2:1–2:14.

[65]  Oracle. *HPROF: A Heap/CPU Profiling Tool*. http://docs.oracle.com/javase/8/docs/technotes/samples/hprof.html. 2016. (Visited on 2022-12-30).

[66]  ej-technologies. *JProfiler*. https://www.ej-technologies.com/products/jprofiler/overview.html. 2021. (Visited on 2022-12-30).

[67]  YourKit GmbH. *YourKit*. https://www.yourkit.com. 2021. (Visited on 2022-12-30).

[68]  Oracle. *JDK Mission Control*. https://www.oracle.com/java/technologies/jdk-mission-control.html. 2022. (Visited on 2022-12-30).

[69]  Sameer Shende and Allen Malony. "The Tau Parallel Performance System". In: *Int. J. High Perform. Comput. Appl.* 20.2 (2006), 287–311. DOI: 10.1177/1094342006064482.

[70]  Intel. *Intel VTune Amplifier*. https://software.intel.com/en-us/intel-vtune-amplifier-xe. 2021. (Visited on 2022-12-30).

[71]  Google. *Class com.google.common.collect.Streams*. https://guava.dev/releases/snapshot-jre/api/docs/com/google/common/collect/Streams.html. 2022. (Visited on 2022-12-30).

[72]  jOOQ. *jOOQ/jOOL*. https://github.com/jOOQ/jOOL. 2020. (Visited on 2022-12-30).

[73]  Tagir Valeev. *StreamEx*. https://github.com/amaembo/streamex. 2022. (Visited on 2022-12-30).

[74]  Paul Chiusano and Rnar Bjarnason. *Functional Programming in Scala*. 1st. USA: Manning Publications Co., 2014. ISBN: 1617290653.

[75]  Apache Groovy project. *Package java.util.stream*. https://docs.groovy-lang.org/latest/html/groovy-jdk/java/util/stream/package-summary.html. 2022. (Visited on 2022-12-30).

[76]  Kotlin Foundation. *Package kotlin.streams*. https://kotlinlang.org/api/latest/jvm/stdlib/kotlin.streams/. 2022. (Visited on 2022-12-30).

[77]  Eduardo Rosales, Matteo Basso, Andrea Rosà, Mariano Marciello, and Walter, Binder. "Profiling and Optimizing Java Streams (artifact)". In: *<Programming>*. 2023. DOI: 10.5281/zenodo.7557341.
10:38

**About the authors**

**Eduardo Rosales** is a Postdoctoral Researcher in the Faculty of Informatics of Università della Svizzera italiana (USI), Switzerland. He holds an MSc from Universidad de los Andes, Colombia. His research interests include cluster/grid/cloud computing, program analysis, and concurrent and parallel programming. Contact him at rosale@usi.ch

**Matteo Basso** is a PhD student in the Faculty of Informatics of Università della Svizzera italiana (USI), Switzerland. He received a joint MSc from Università della Svizzera italiana (USI), Switzerland, and Università degli Studi di Milano-Bicocca, Italy. His research interests include compilers, managed languages and runtimes, and code parallelization. Contact him at matteo.basso@usi.ch

**Andrea Rosà** is a Postdoctoral Researcher in the Faculty of Informatics of Università della Svizzera italiana (USI), Switzerland. He holds a PhD from USI, as well as a professorship habilitation and an MSc from Politecnico di Milano, Italy. His research interests include concurrency and parallelism, program analysis, compilers, virtual machines, and big-data computing. Contact him at andrea.rosa@usi.ch

**Walter Binder** is a Full Professor in the Faculty of Informatics of Università della Svizzera italiana (USI), Switzerland. He holds an MSc, a PhD, and a Venia Docendi from TU Wien, Austria. Before joining USI, he was a post-doctoral researcher at the Artificial Intelligence Laboratory, École Polytechnique Fédérale de Lausanne, Switzerland. His main research interests are in the areas of program analysis, concurrent and parallel programming, and virtual machines. Contact him at walter.binder@usi.ch